# Epitaxial NbN-junction-based self-shunted superconducting flux qubit with high anharmonicity

Tokunoshin Uchida[a], Shingo Nozaki[a], Daiki Kurihara[a], Hiroki Kutsuma[a], Shotaro Shirai[b,c], Atsushi Noguchi[b,c,d], Hirotaka Terai[e], Taro Yamashita[a,*]

[a]*Department of Applied Physics, Graduate School of Engineering, Tohoku University, Sendai, 980-8579, Miyagi, Japan*
[b]*RIKEN Center for Quantum Computing, Wako, 351-0198 Saitama, Japan.*
[c]*Komaba Institute for Science (KIS), The University of Tokyo, Meguro-ku, 153-8902, Tokyo, Japan.*
[d]*Inamori Research Institute for Science (InaRIS), Kyoto-shi, 600-8411 Kyoto, Japan.*
[e]*Advanced ICT Research Institute, National Institute of Information and Communications Technology (NICT), Kobe, 651-2492 Hyogo, Japan.*

**Abstract**

We report a demonstration of a self-shunted superconducting flux qubit (SSFQ) as an ultracompact, strongly anharmonic, and long-lifetime qubit by using the full epitaxially grown NbN/AlN/NbN Josephson junctions. The NbN-junction-based SSFQ achieved a high anharmonicity of 797 MHz as well as drastically reduced footprint of the qubit compared to capacitively shunted (C-shunt) flux qubits with a large shunt capacitor. The lifetime of the SSFQ reached 13 – 25 μs, which is comparable to or relatively higher than that of NbN-based C-shunt flux qubits and Al-based flux qubits with more than 10 times smaller junctions. This indicates that the large epitaxial NbN-based junction can be adopted for SSFQ as well as merged-element transmon with the relatively large junctions. The results presented in this work will offer a new choice of the scalable qubits as well as the material platform toward the realization of large-scale superconducting quantum computers.

## 1. Introduction

Recently, research and development on superconducting quantum computers has progressed rapidly. The energy relaxation time (lifetime) of transmon qubits, as the current leading architecture, has reached up to 1 ms [1, 2]. However, for practical applications, achieving large-scale integration is essential, and transmons face key challenges. The first challenge is the shunt capacitor added in parallel to the Josephson junction to reduce charge noise. Because the shunt capacitor occupies a large footprint of several hundred micrometers, it becomes a major bottleneck for high-density integration. To overcome this limitation, the merged-element transmon (MET or mergemon) was proposed [3-6]. By expanding the junction area relative to standard transmons to achieve a junction capacitance on the order of tens of femtofarads, the MET eliminates the external shunt capacitor, thereby improving the integration density. So far, the lifetime exceeding 200 μs have been reported for METs [4].

The second challenge in transmons is their low anharmonicity (typically 200–400 MHz), which limits gate operation speed and leads to frequency crowding in large-scale quantum circuits. In contrast, flux qubits possess much larger anharmonicities [7-12]. Recently, capacitively shunted (C-shunt) flux qubits, which incorporate an additional shunt capacitor, have demonstrated long lifetimes of 55 μs (with 2D resonator) [8] and 90 μs (with 3D cavity) [9]. Nonetheless, like transmons, the large device footprint caused by the shunt capacitor poses a barrier to integration. To address this issue, the self-shunted superconducting flux qubit (SSFQ) has been proposed, and systematic numerical simulations have been performed [13]. As a result, it has been demonstrated theoretically that a long lifetime and high anharmonicity could be simultaneously achieved by increasing the junction size (like MET) without the external shunt capacitor.

In both SSFQs and METs, energy losses in/around the junction will be crucial for the lifetime because electromagnetic fields are concentrated near the junction [3]. In conventional Al-based junctions, the amorphous $AlO_x$ barrier has a high density of two-level systems (TLSs) that act as loss sources [14,15] and may degrade the performance of SSFQs and METs. On the other hand, fully epitaxial NbN/AlN/NbN junction qubits have been demonstrated in recent years, where the epitaxial AlN barrier is expected to suppress TLS density [10, 16-21]. So far, a lifetime of over 10 μs has been reported in NbN-based C-shunt flux qubits [10, 20], and epitaxially grown junction technology is expected to be well suited for SSFQs and METs. Furthermore, NbN-based qubits are advantageous for large-scale integration because their fabrication process is compatible with standard semiconductor processing and does not require liftoff or shadow-angle

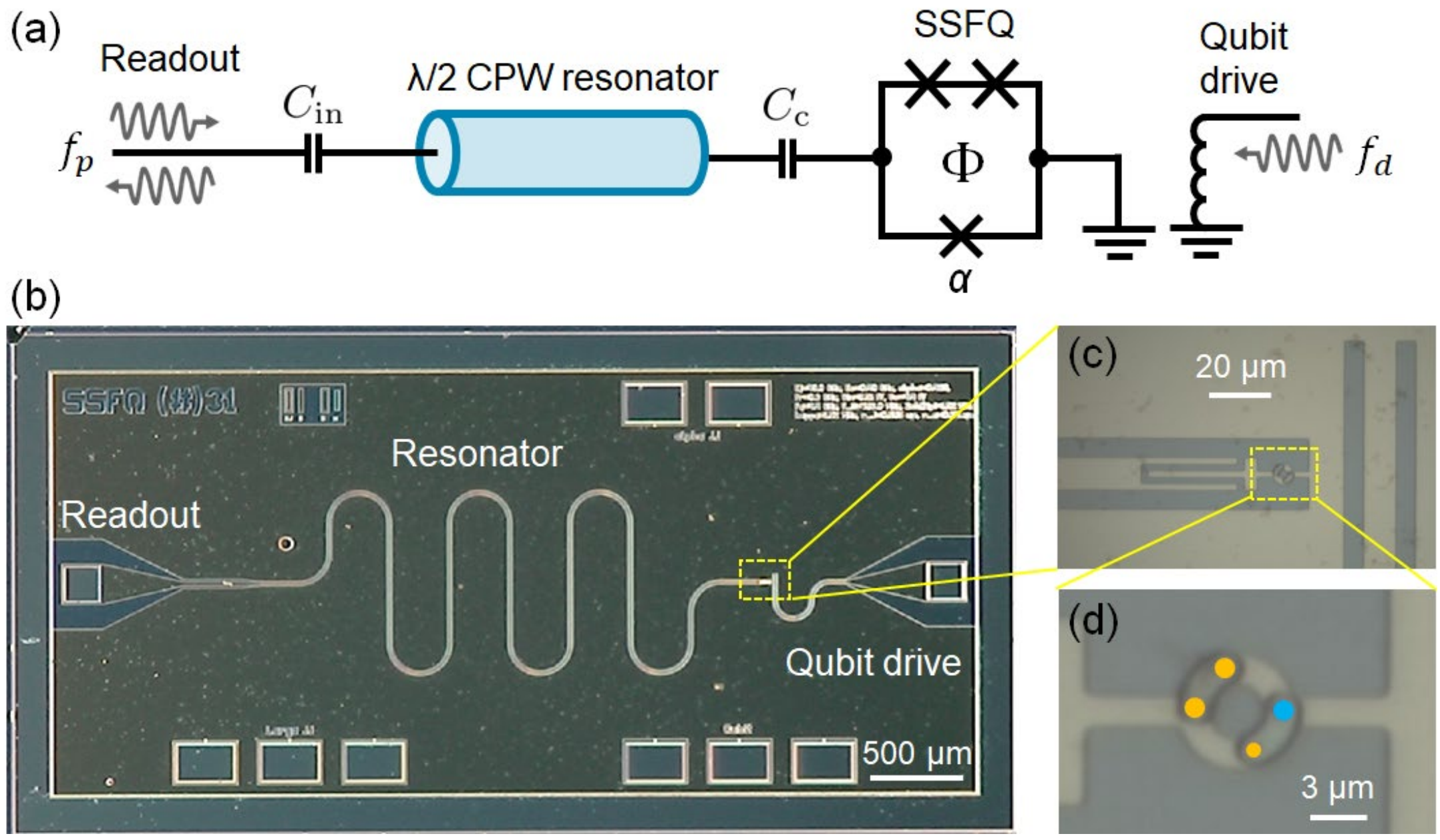

Fig. 1. (a) Schematic diagram of the device. (b) Optical microscope image of the device. The chip size is 2.5 × 5 mm$^2$. (c) Image of the device with coupling capacitor magnified for the SSFQ part. (d) False-color image of the SSFQ with three Josephson junctions. Josephson junction and via-hole are marked by orange and blue circles, respectively.

evaporation, as typically used for Al-based qubits.

In this study, we present the experimental demonstration of an SSFQ fabricated with fully epitaxial NbN-based junctions. By designing and fabricating an SSFQ by relatively large junctions with a diameter of around 1 μm, we successfully achieved both a long lifetime of 25 μs and a large anharmonicity of 797 MHz.

## 2. Device design and fabrication

First, based on theoretical calculations, we designed an SSFQ consisting of three Josephson junctions (one of which has a junction area smaller than the other two by a factor of $\alpha$) to achieve a high anharmonicity while suppressing energy relaxation induced by charge and flux noise [13]. The design parameters are as follows: $E_J/h = 49.8$ GHz, $E_C/h = 0.480$ GHz, $\alpha = 0.429$, $f_{01} = 4.404$ GHz, $\mathcal{A} = 760$ MHz where $E_J$ and $E_C$ represent the Josephson energy and charging energy of the larger junctions, respectively, $f_{01}$ is the qubit frequency, and $\mathcal{A}$ is the anharmonicity at the flux-insensitive point. In this SSFQ, by setting the area of the larger junctions to approximately 1 μm$^2$, we secured a total capacitance of 40 – 50 fF, which is comparable to that of a C-shunt flux qubit, thereby expecting a suppression of the charge noise. Figure 1(a) shows a schematic circuit diagram. The SSFQ is coupled to a half-wavelength ($\lambda$/2) coplanar waveguide (CPW) resonator for readout and a drive line for qubit control. The resonance frequency of the readout resonator ($E_r/h$), $C_{\mathrm{in}}$, and $C_{\mathrm{c}}$ were designed to be 8.3 GHz, 6.5 fF, and 4.4 fF, respectively. The detail of the Hamiltonian is described in Appendix A.

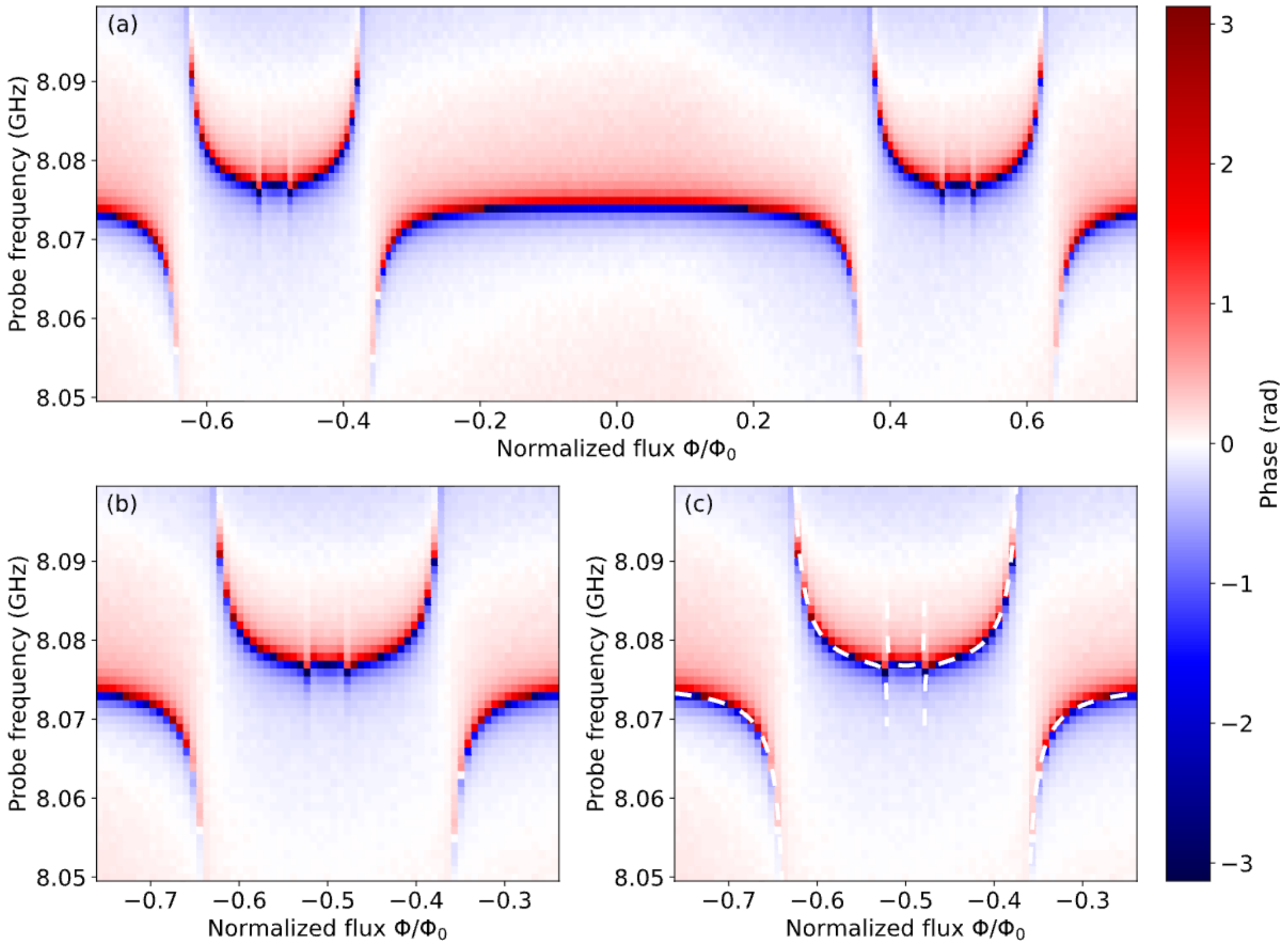

Fig. 2. Spectroscopy of the coupled system with the SSFQ and the resonator. (a) Phase of the reflection coefficient $S_{11}$ as a function of the probe frequency and the normalized flux bias. (b) The same data as (a) magnified around flux-insensitive point. (c) The same data as (b) with the fitting results overlaid by the white dashed curves.

Figure 1(b) presents a photograph of the fabricated device. Device fabrication was performed on a high-resistivity Si (100) substrate. After depositing a 100-nm-thick TiN buffer layer, a NbN/AlN/NbN trilayer epitaxial film (total thickness of 200 nm) was deposited to form the junctions. The critical current density $J_c$ of the junction can be controlled via the deposition time of the AlN layer, and measurements from a test chip fabricated on the same substrate yielded 12 – 14 A/cm$^2$, which is close to the designed value of 10 A/cm$^2$. The CPW resonator and ground plane were fabricated from TiN, while NbTiN was utilized for the top wiring layer. Here, NbTiN was adopted for the gap engineering because the superconducting gap of NbTiN is smaller than that of NbN [22]. Details regarding the fabrication process are shown in Appendix B. As shown in Figs. 1(c) and (d), the SSFQ requires no shunt capacitor to reduce charge noise, and thus the footprint of the qubit is drastically reduced to approximately 1/3100 (1/200 even including the coupling capacitor) that of C-shunt flux qubits [10].

## 3. Experimental results and discussion

The device was mounted in a gold-plated sample holder, thermally anchored to the mixing chamber stage of a dilution refrigerator, and cooled to below 13 mK. The sample holder was enclosed in a three-layer magnetic shield (one superconducting Al layer and two μ-metal layers), with an additional μ-metal shield installed inside the room-temperature shield. A superconducting coil was wound around the sample holder, allowing a magnetic flux bias $\Phi$ to be applied to the device using a room-temperature current source (Yokogawa GS200). For spectroscopy measurements, a vector network analyzer (Keysight M9805A) and a signal generator (AnaPico APMS12G-2) were used. The coherence properties of the qubit were evaluated using a commercial Quantum Control System (Keysight QCS). Details of the measurement setup are shown in Appendix C.

Figure 2 shows the spectroscopy measurements of the readout resonator. Figure 2(a) plots the phase of the reflection coefficient $S_{11}$ as a function of the probe frequency $f_p$ and normalized flux bias $\Phi/\Phi_0$ ($\Phi_0$ is the magnetic flux quantum). As shown in Fig. 2(a), a periodic anti-crossing indicating the presence of SSFQ and its coupling to the resonator was observed. At the flux-insensitive point ($\Phi/\Phi_0 = \pm 0.5$), the resonance frequency was 8.0772 GHz. Furthermore, as shown in Figs. 2(a) and (b), other distinct anti-crossings were observed in the vicinity of the flux-insensitive point ($\Phi/\Phi_0 \approx \pm 0.48$ and $\pm 0.52$). These anti-crossings originate from the coupling between the resonator and qubit in the higher-energy states (discussed below). Figure 2(c) is the same as (b), but the fitting result by the Hamiltonian is overlaid by the dashed curves (details in Appendix A). By using the equation of the reflection coefficient [23], we extracted the loaded, internal, and external quality factors $Q_L$, $Q_i$, and $Q_e$, of the readout resonator, and obtained to be 4.319 × 10$^3$, 1.735 × 10$^4$, and 5.735 × 10$^3$ at the flux-insensitive point, respectively. In the present sample, the loss tangent of the TiN resonator on the Si substrate is $\tan\delta = 1/Q_i =$ 5.764 × 10$^{-5}$, which is larger than typical values of NbN and TiN resonators [10, 24]. Although the reason has not been fully understood, one possible cause is the damage of the TiN and/or Si surfaces by RF sputtering of the $SiO_2$ as an interlayer dielectric layer.

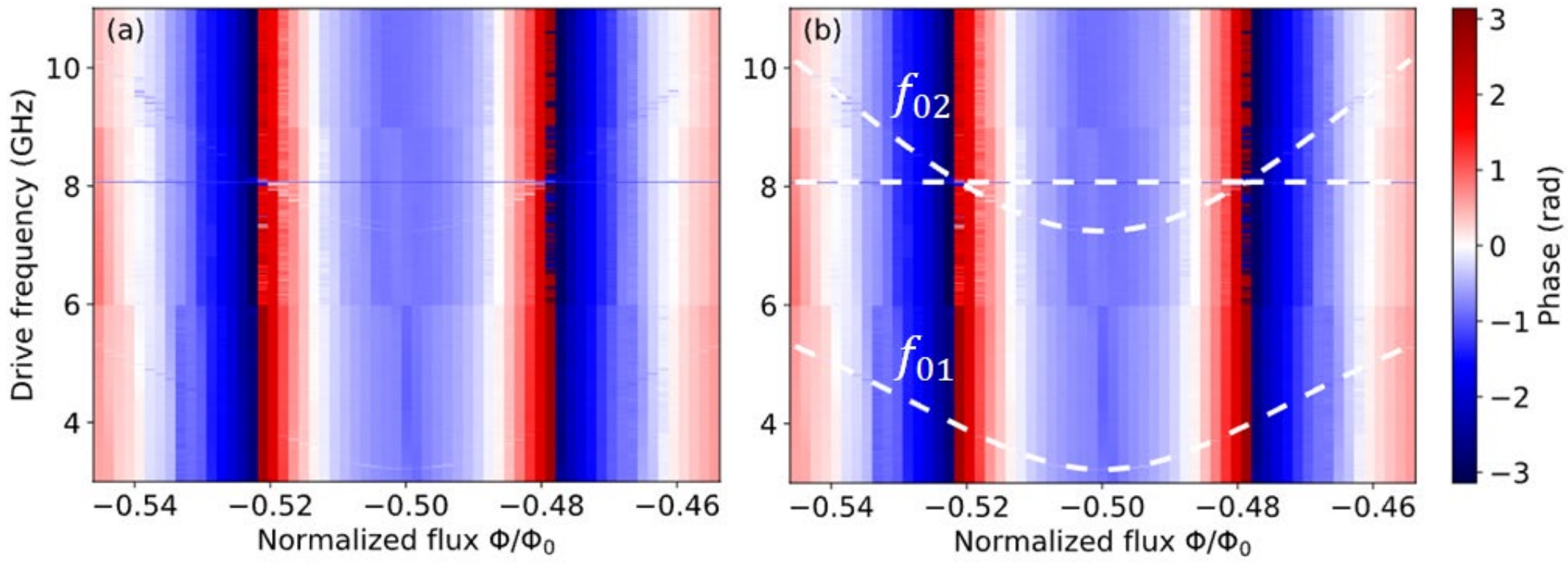

Fig. 3. Spectroscopy of the coupled system with the SSFQ and the resonator. (a) Phase of the reflection coefficient $S_{11}$ as a function of the drive frequency for the qubit drive port and the normalized flux bias. The probe frequency was fixed to be the resonance frequency of the readout resonator at the flux-insensitive point. Drive power at the output of the microwave source was set to be –30 dBm, –10 dBm, and 0 dBm for the drive frequency ranges of 3 – 6 GHz, 6 – 9 GHz, and 9 – 11 GHz, respectively. (c) The same data as (a) with the fitting results overlaid by the white dashed curves.

Figure 3(a) plots the phase of $S_{11}$ as a function of the drive frequency of the continuous-wave microwave input to the qubit drive port $f_d$ and $\Phi/\Phi_0$, with the probe frequency fixed at $f_p = f_r$ (where $f_r$ is the resonance frequency of the readout resonator at the flux-insensitive point). A flux-dependent qubit frequency was observed, yielding $f_{01} = 3.2255$ GHz at the flux-insensitive point. To experimentally extract the device parameters of the SSFQ, we fitted the obtained spectrum to the theoretical curves calculated by diagonalizing the Hamiltonian (details in Appendix A) as shown in Fig. 3(b). As shown in the figure, the experimental data were well fitted for the transition frequencies $f_{01}$ ($f_{02}$), corresponding to the energy differences between the ground state and the first (second) excited states. From the fitted result, we confirmed that the anti-crossing near the flux-insensitive point seen in Fig. 2 is caused by coupling between the second excited level of the qubit ($f_{02}$) and the resonator. Table 1 compares the extracted device parameters obtained from the numerical fit with the design parameters. Although $E_C$, $C_C$, and $E_r$ was lower than the design values, we obtained device parameters close to the target values for $E_J$ and $\alpha$. Table 2 shows a summary of the designed and measured values of $f_r$, $f_{01}$, the coupling strength $g/2\pi$, the dispersive shift $2\chi/2\pi$, and the anharmonicity $\mathcal{A}$ at the flux-insensitive point. The measured values of $g/2\pi$ and $2\chi/2\pi$ were estimated by the fitting results. The anharmonicity, $\mathcal{A}$, a crucial parameter for the SSFQ, reached a large value of 797 MHz, which is close to the designed value and more than double that of a typical transmon.

Table 1. Parameters of the device.

| | $E_J/h$ (GHz) | $E_C/h$ (GHz) | $\alpha$ | $C_c$ (fF) | $E_r/h$ (GHz) |
|---|---|---|---|---|---|
| Designed | 49.8 | 0.480 | 0.429 | 4.40 | 8.3000 |
| Fitting parameters | 47.7 | 0.359 | 0.465 | 3.09 | 8.0957 |

Table 2. Parameters of the device for the flux bias at the flux-insensitive point.

| | $f_r$ (GHz) | $f_{01}$ (GHz) | $g/2\pi$ (MHz) | $2\chi/2\pi$ (MHz) | $\mathcal{A}$ (MHz) |
|---|---|---|---|---|---|
| Designed | 8.2742 | 4.4040 | 83.17 | 1.49 | 760 |
| Measured | 8.0772 | 3.2255 | 42.54 | 0.307 | 797 |

Next, to evaluate the coherence properties of the SSFQ, we measured the lifetime ($T_1$), the Ramsey decay time ($T_2^*$), and the Hahn-echo coherence time ($T_{2E}$). The corresponding control pulse sequences are shown in Figs. 4(a), 5(a), and 5(b), respectively. Figure 4(a) shows the energy relaxation measurement along with the exponential decay fit by using $\exp(-\tau/T_1)$, yielding $T_1 = 24.7$ μs which is comparable to the highest values reported for the flux qubit without shunt capacitor [11, 12]. To investigate the fluctuations in $T_1$, we performed 188 repeated measurements over 25 hours. As a result, as shown in Fig. 4(b), the measured $T_1$ fluctuated within a range of 13 – 25 μs, with a center value of 15.5 μs and a standard deviation of 0.15 by a log-normal distribution fit. These values are comparable to those reported for the NbN-based C-shunt flux qubits while the maximum $T_1$ was higher. In the NbN-based C-shunt flux qubit on the Si substrate, the obtained $T_1$ was consistent with the value estimated from the internal quality factor of the resonator and the participation ratio of the shunt capacitor [10]. On the other hand, SSFQ reached similar values of $T_1$ despite the internal quality factor being one order of magnitude lower. This may be attributed to the absence of a shunt capacitor in the SSFQ, which concentrates the electric field around the junction and thereby reduces its susceptibility to dielectric loss in the substrate. As a comparison to Al-based flux qubits without a shunt capacitor [11, 12], the lifetime of our NbN-based SSFQ is comparable to their best values despite the junction size of the SSFQ is more than 10 times larger than that of Al-based qubits [11]. To achieve the longer lifetime reported for the C-shunt flux qubits [8, 9] by the SSFQ, further improvements in the qubit design and fabrication will be required. Figures 5(a) and (b) show phase relaxation from Ramsey and Hahn-echo experiments, respectively. We obtained $T_2^* = 0.52$ μs and $T_{2E} = 3.4$ μs from the fitting to data. By canceling the low-frequency noise in the Hahn-echo measurement, the coherence time was improved by more than a factor of six. Pure Hahn-echo dephasing time is estimated to be around 3.8 μs and coherence time was mainly limited by the pure dephasing.

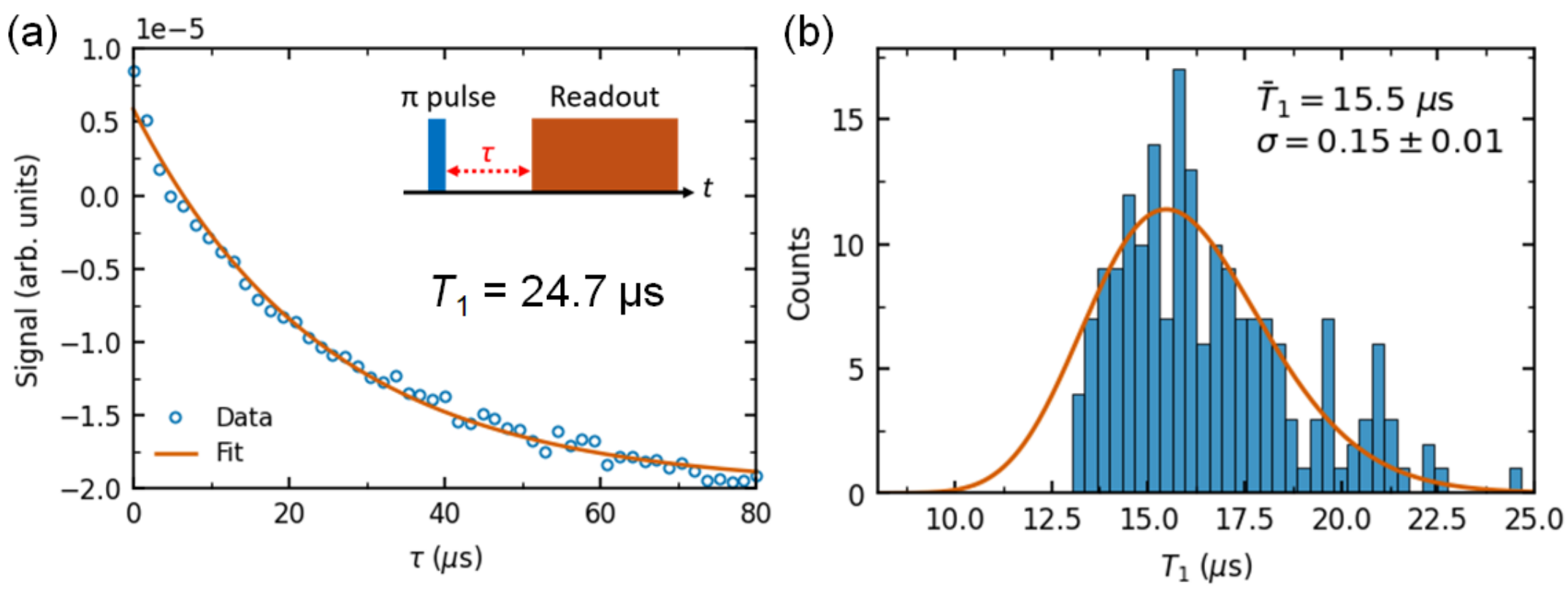

Fig. 4. (a) $T_1$ decay profile with an exponential fit (solid line) with $T_1 = 24.7$ μs. The inset shows the pulse sequence of the $T_1$ measurement. (b) Histogram of the $T_1$ values with a log-normal distribution fit (solid line) with center value $\bar{T}_1 = 15.5$ μs and standard deviation $\sigma = 0.15$.

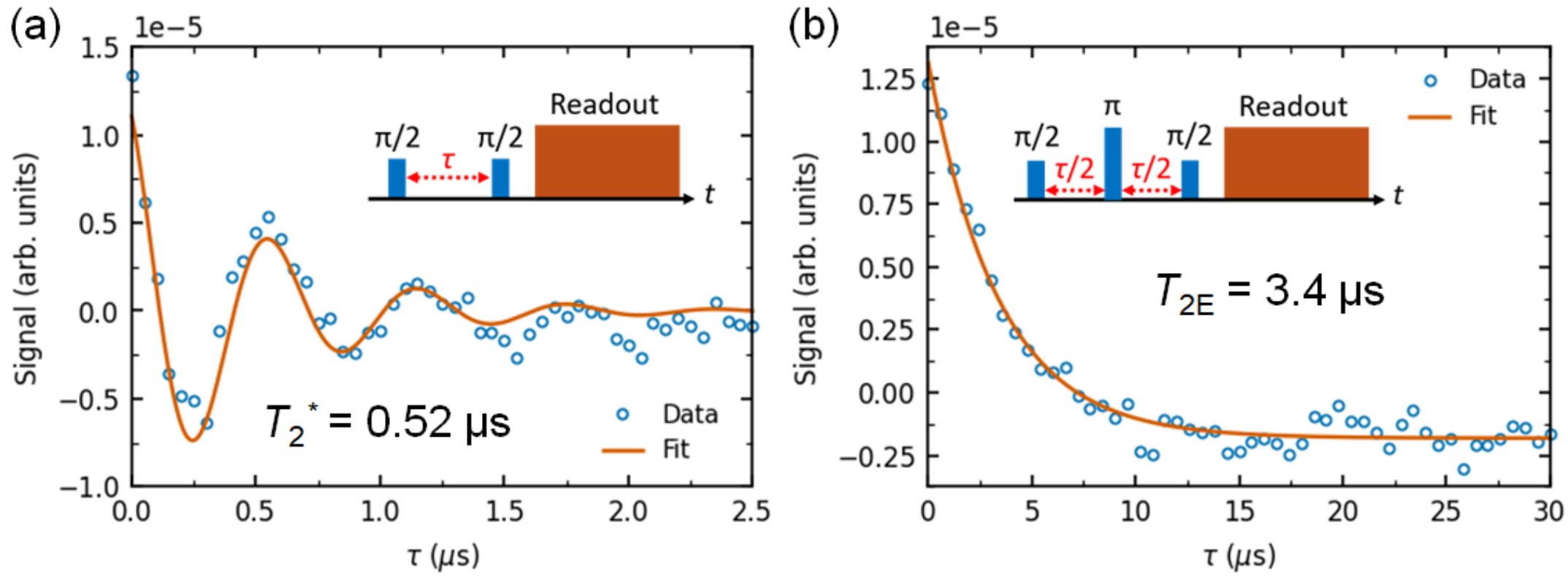

Fig. 5. (a) Ramsey fringe signal with $T_2^* = 0.52$ μs. (b) Hahn-echo signal with $T_{2E} = 3.4$ μs. The solid lines show the fitting curve. The pulse sequences of the measurements are shown in the insets.

## 4. Conclusion

In conclusion, we demonstrated a self-shunted superconducting flux qubit (SSFQ) as an ultracompact, strongly anharmonic, and long-lifetime qubit by using the fully epitaxial NbN-based Josephson junctions. The NbN-based SSFQ achieved a high anharmonicity of 797 MHz as well as drastically reduced footprint of the qubit part compared to the capacitively shunted (C-shunt) flux qubits, which is crucial for the large-scale integration of superconducting qubits. The lifetime of the SSFQ was obtained to be 13 – 25 μs comparable to or relatively higher than that of NbN-based C-shunt flux qubits, even without an additional shunt capacitor. Furthermore, the obtained SSFQ lifetime is also comparable to the best values of the Al-based flux qubit with more than 10 times smaller junctions. This indicates that the large epitaxial NbN-based junction is suitable for SSFQs as well as merged-element transmons. The results presented in this work will offer a new choice of the scalable qubits as well as the material platform toward the realization of large-scale superconducting quantum computers.

## CRediT authorship contribution statement

**Tokunoshin Uchida**: Data curation, Formal analysis, Funding acquisition, Investigation, Visualization, Writing – review & editing. **Shingo Nozaki**: Data curation, Formal analysis, Investigation, Visualization, Writing – review & editing. **Daiki Kurihara**: Data curation, Formal analysis, Visualization, Writing – review & editing. **Hiroki Kutsuma**: Investigation, Writing – review & editing. **Shotaro Shirai**: Resources, Writing – review & editing. **Atsushi Noguchi**: Resources, Writing – review & editing. **Hirotaka Terai**: Data curation, Investigation, Writing – review & editing. **Taro Yamashita**: Conceptualization, Formal analysis, Funding acquisition, Resources, Supervision, Visualization, Writing – original draft.

## Declaration of Competing Interest

The authors declare that they have no known competing financial interests or personal relationships that could have appeared to influence the work reported in this paper.

**Acknowledgements**
The authors would like to thank K. Takaki and M. Setoura for their assistance with sample fabrication, and K. Inomata and T. Yamamoto for fruitful discussion. This work was partly supported by Grant-in-Aid for JSPS Fellows (25KJ0535), JST CREST (JPMJCR24I5), Grant-in-Aid for Scientific Research (S) (25H00406), JST Moonshot R&D Program (JPMJMS2067, JPMJMS256L). T.Y. and H.K. acknowledge Center for Heterogeneous Quantum/Material Fusion Technologies, Center for Key Interdisciplinary Research, Tohoku University.

**Appendix A. Full Hamiltonian of SSFQ coupled with readout resonator**

We used the following Hamiltonian $\hat{\mathcal{H}}$ which consists of qubit term $\hat{\mathcal{H}}_\mathrm{q}$, resonator term $\hat{\mathcal{H}}_\mathrm{r}$, and coupling term $\hat{\mathcal{H}}_\mathrm{c}$ for designing the sample and fitting the experimental data:

$$\hat{\mathcal{H}} = \hat{\mathcal{H}}_\mathrm{q} + \hat{\mathcal{H}}_\mathrm{r} + \hat{\mathcal{H}}_\mathrm{c}, \quad (\mathrm{A.1})$$

where

$$\hat{\mathcal{H}}_\mathrm{q} = 4E_\mathrm{C}\frac{(1+\alpha)(1+\gamma)+\zeta}{(1+2\alpha)(1+\gamma)+2\zeta}(\hat{n}_1^2+\hat{n}_2^2) + 8E_\mathrm{C}\frac{\alpha(1+\gamma)+\zeta}{(1+2\alpha)(1+\gamma)+2\zeta}\hat{n}_1\hat{n}_2 - E_\mathrm{J}\cos\hat{\varphi}_1 - E_\mathrm{J}\cos\hat{\varphi}_2 - \alpha E_\mathrm{J}\cos\left(\hat{\varphi}_1-\hat{\varphi}_2+2\pi\frac{\Phi}{\Phi_0}\right), \quad (\mathrm{A.2})$$

$$\hat{\mathcal{H}}_\mathrm{r} = f_\mathrm{r}\sqrt{\frac{1+2\alpha+2\zeta}{(1+2\alpha)(1+\gamma)+2\zeta}}\left(\hat{a}^\dagger\hat{a}+\frac{1}{2}\right), \quad (\mathrm{A.3})$$

$$\hat{\mathcal{H}}_\mathrm{c} = -2i\frac{\sqrt{\gamma\zeta f_\mathrm{r}E_\mathrm{C}}}{\sqrt[4]{(1+2\alpha+2\zeta)[(1+2\alpha)(1+\gamma)+2\zeta]^3}}(\hat{n}_1-\hat{n}_2)(\hat{a}^\dagger-\hat{a}). \quad (\mathrm{A.4})$$

Here $\hat{n}_i$ and $\hat{\varphi}_i$ ($i = 1, 2$) are the Cooper pair number operator and the phase difference operator of the $i$-th junction, respectively, and $\hat{a}^\dagger(\hat{a})$ is the creation (annihilation) operator. $E_J$ and $E_C$ are the Josephson energy and charging energy of the larger junctions, respectively, and $f_r$ is the resonator frequency. $\alpha$ is the junction area ratio between the smaller and larger junctions. The ratio $\gamma = C_\mathrm{c}/C_\mathrm{r}$, $\zeta = C_\mathrm{c}/C_\mathrm{J}$ with the coupling capacitance $C_\mathrm{c}$, the resonator capacitance $C_r$, and the junction capacitance $C_\mathrm{J}$. $\Phi$ and $\Phi_0$ are the external magnetic flux and the superconducting flux quantum, respectively.

**Appendix B. Device fabrication**

Samples were fabricated by using epitaxial growth of NbN/AlN/NbN junctions on high resistivity Si (100) wafer with TiN buffer. The fabrication flow is as follows: (i) The NbN (150 nm)/AlN (~2 nm)/NbN (50 nm) tri-layers are epitaxially grown on a high-resistivity (> 20 kΩ·cm) Si (100) substrate with a TiN (100 nm) buffer layer by a dc magnetron sputtering system. (ii) The junction patterns are made by electron beam lithography (ELIONIX ELS-F125) using a ZEP-520A positive resist. A thin MgO and NbN layers are deposited, followed by a lifted-off in NMP solution. (iii) These thin layers serve as an etching mask for the junction definition and NbN/AlN/NbN tri-layers are etched by reactive ion etching (RIE) with a $CF_4$ plasma. The mask layers are eliminated by an etching subsequently after the junction RIE process. (iv) Then the resonator and base patterns are made by maskless lithography (Heidelberg MLA150) using AZ703 photoresist, and the TiN layer were etched by RIE with Ar and $CF_4$ plasma. (v) Next a $SiO_2$ film that serves as an isolating layer between the base and wiring layers of the qubit was deposited. (vi) A planarization process by CMP is carried out to achieve a stable contact hole between the submicron-size junctions and the wiring layer. (vii) After forming contact holes between the base electrode and the wiring layer, the 300 nm-thick NbTiN wiring layer is deposited by dc magnetron sputtering and patterned. (viii) Finally, the $SiO_2$ isolation layer, which may contain unwanted TLSs, was removed by etching with buffered hydrogen fluoride (BHF) before measurement.

## Appendix C. Measurement setup

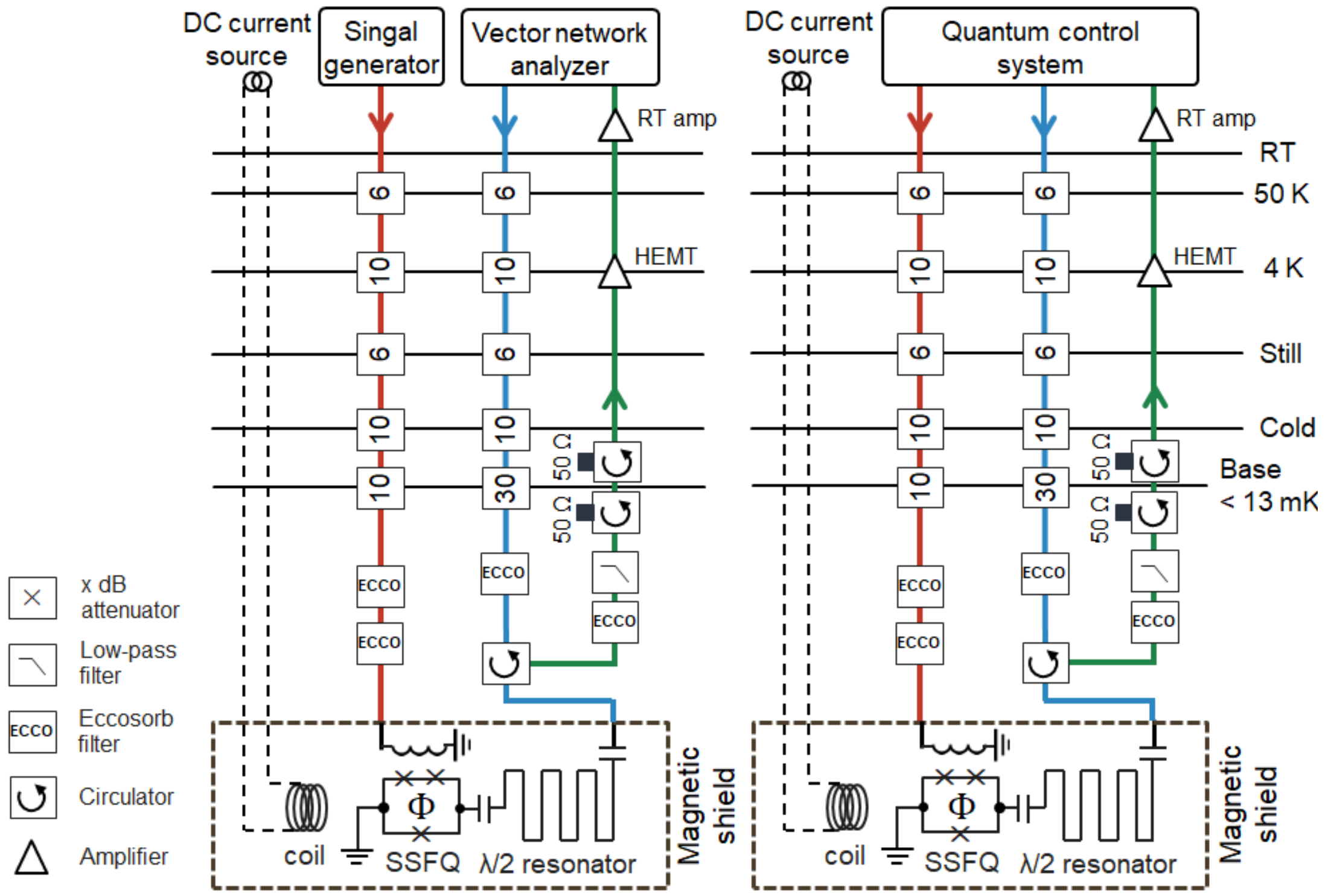

Fig. A.1. Experimental setup diagrams for (left) spectroscopy measurements and (right) qubit coherence measurements.

Figure A.1 shows an experimental setup diagram for the measurements. Qubit control and readout microwave signals are generated by a signal generator (AnaPico APMS12G-2) and vector network analyzer (Keysight M9805A) for frequency domain measurements (Figs. 2 and 3) as shown in left panel of Fig. A.1. The microwave signals are sent through an input microwave line with attenuators of 42 dB in total and eccosorb filters for the qubit control (red line). For the readout line, the microwave pulses are further attenuated by 20 dB (blue line) and are routed to the resonator through a circulator to separate the input and output (reflected) waves.

The output line for readout (green line) includes an eccosorb filter, a low-pass filter (cutoff frequency: 12.4 GHz), and two isolators (4–12 GHz circulators with 50 Ω) to prevent noise/blackbody radiation coming from the cryogenic HEMT amplifier. The reflected signal is amplified by the HEMT amplifier and a room-temperature amplifier. For the time-domain measurements (Figs. 4 and 5), we replace the readout and the control circuits with a Quantum Control System (QCS, Keysight) as shown in right panel of Fig. A.1.

The external flux in the flux qubit is controlled by the local flux bias Φ which is generated by a superconducting coil wound around the sample package mounted inside the cryogenic magnetic shield.